\documentclass[12pt,preprint,epsf,epsfig,graphics]{aastex}

\shorttitle{Planet-Induced Modulation of Ca II H \& K}
\shortauthors{Shkolnik, Walker, \& Bohlender}

\begin{document}


\title{Evidence for Planet-induced Chromospheric Activity on HD~179949}


\author{E. Shkolnik\altaffilmark{1} and G.A.H. Walker\altaffilmark{1}}
\affil{Department of Physics \& Astronomy, University of British Columbia,
    6224 Agricultural Rd., Vancouver  BC, Canada V6T 1Z1}
\email{shkolnik@astro.ubc.ca}
\email{walker@astro.ubc.ca}

\altaffiltext{1}{Visiting Astronomer, Canada-France-Hawaii Telescope,
operated by the National Research Council of Canada, the Centre
National de la Recherche Scientifique of France, and the University of
Hawaii.}

\and

\author{D.A. Bohlender\altaffilmark{1}}
\affil{Herzberg Institute for Astrophysics, National Research Council of Canada\\ 
Victoria BC, Canada V9E 2E7}
\email{david.bohlender@nrc-cnrc.gc.ca}

\begin{abstract}
We have detected the synchronous enhancement of Ca~II H \& K emission with
the short-period planetary orbit in HD~179949.  High-resolution spectra
taken on three observing runs extending more than a year show the
enhancement coincides with $\phi \sim$ 0 (the sub-planetary point) of the
3.093-day orbit with the effect persisting for more than 100 orbits. The
synchronous enhancement is consistent with planet-induced chromospheric
heating by magnetic rather than tidal interaction. Something which can only
be confirmed by further observations. Independent observations are needed to
determine whether the stellar rotation is sychronous with the planet's
orbit. Of the five 51~Peg-type systems monitored, HD~179949 shows the
greatest chromospheric H \& K activity. Three others show significant nightly
variations but the lack of any phase coherence prevents us saying whether
the activity is induced by the planet. Our two standards, $\tau$ Ceti and
the Sun, show no such nightly variations.

\end{abstract}




\section{Introduction}

Current planet detection methods provide basic information: a minimum mass if the orbital inclination is not known and an albedo-dependent estimate of surface temperature. In the case of the transiting system, HD~209458, the planet's density is also determined with a possible atmospheric sodium detection (Charbonneau 2002) and excess Lyman $\alpha$ absorption during transit (Vidal-Madjar et al. 2003).  There still remains a lack of constraints on extrasolar planetary structure leaving astronomers to explore new observational probes.

Cuntz, Saar \& Musielak (2000) suggested that there may be an observable interaction between a parent star and a close-in giant planet, specifically an external heating of the star's outer atmosphere.  The effect could be tidal, magnetic, or a combination of the two.  In the case of tidal heating, the acoustic wave energy thought to contribute to the heating of the upper atmosphere has a dependence on the local turbulent velocity ($v_{t}$) as  $v_{t}^8$ (Muzielak et al. 1994).  The steepening of the density gradient in the two tidal bulges would
increase the turbulent velocity producing excess shocks, waves,
and flows in the upper atmosphere of the star.  Magnetic wave energy has a
$v_{t}^6$ dependence (Ulmschneider \& Muzielak 1998) such that an
increase in $v_{t}$ would enhance local dynamo generation and hence the local
magnetic energy will also increase. Thus, even small increases in $v_{t}$  would cause a dramatic intensification in nonradiative heating. If planet-induced heating is confined to a narrow range in stellar longitude, the heated regions would track the planet. This implies that the period of any observed activity would be correlated with the planet's orbit such that tidally induced activity has a period of $\sim P_{orb}/2$ and magnetic activity, a period of $\sim P_{orb}$.  It is also possible that the increased dynamo action would be confined to the turbulent layer below the convection zone (the tachocline). In this case, any dynamo action would be spread out over a wide range of longitudes and  the excess heating would not show strong periodic variation (Saar \& Cuntz  2001). 

Since the tidal and magnetic interaction depend on the distance from the planet to the star as $1/d^{3}$ and $1/d^{2}$, respectively, it is best to look at the tightest systems.  Of the $\sim$~100 known extrasolar planets, 16\% have semi-major axes of less than 0.1 AU and masses comparable to Jupiter's (Schneider 2003). It is expected that these planets have magnetic fields similar to Jupiter's (4.3 G). However if tidally locked, the planets' fields might be substantially smaller.  It is also reasonable to assume that the magnetic interaction would be greatest in the outermost layers of the star, namely the chromosphere, transition region and the corona due to their proximity to the planets and their nonradiative heat sources.  Our program stars have orbital periods between 3.1 and 4.6 days, eccentricities $\sim$ 0 and semi-major axes $< 0.06$ AU.  These systems offer the best chance of observing upper atmospheric heating.  The five systems we chose were $\tau$~Boo, HD~179949, HD~209458, 51~Peg and  $\upsilon$~And and their planetary system parameters are listed in Table 1.  

\section{Chromospheres and Ca II H \& K Emission}

Temperatures in the chromosphere can reach 20,000 K over many scale heights implying that the temperature gradients are not very high. This is due to the large opacity of resonance lines such as H~I, Ca~II, Mg~II and the Lyman continuum which then in the low-density gas radiate and cool the chromosphere. Most chromospheric indicators are UV emission lines not easily accessible from ground-based telescopes.  However, the optical resonance lines of Ca II H  \& K at 3933 \AA\/ and 3968 \AA\/ exhibit chromospheric emission in late-type stars (later than F0) and are sensitive probes of temperature and electron density. 
An example of HD~179949 Ca~II H \& K reversals are shown in Figure 1.  
The double-peaked emission is formed in the chromosphere at T $\sim$ 8000~K where the source function is essentially monochromatic. The central emission is produced at a temperature of $\sim$ 20,000~K (Montes et al. 1994) which appears as a self-reversal due to the large optical depth of the line.  The asymmetry of the emission in late-type dwarfs is thought to be a consequence of upflows in the chromosphere's plage-like regions.  (See Linsky 1980 for a comprehensive observational and theoretical review.)

The broad, deep photospheric absorption and the low level background continuum allow the reversal to be seen at higher contrast. Because of these and the accessibility of the lines from the ground, the Ca~II H \& K reversals are an optimal choice with which to monitor chromospheric heating for the sun-like stars known to host short-period giants planets.  
Saar and Cuntz (2001) looked at the Ca~II~IR line at 8662~\AA, an interlocking line with the Ca~II H \& K resonance lines, with spectra of resolution R = 50,000 and S/N typically of 200.  They looked at five stars with close-in giant planets (including $\tau$~Boo, $\upsilon$~And, and 51~Peg) by simulating the IR equivalent to the Mt. Wilson $S_{HK}$ index and found nothing at the 3 to 5\% level.  

\section{Observations}

High-resolution (R = 110,000) and high signal-to-noise (S/N $\approx$ 500 
per pixel in the continuum and 150 in the H \& K cores) spectra were 
obtained on 3 observing runs at the Canada-France-Hawaii Telescope (CFHT) with the fiber-fed coud\'{e} spectrograph, $Gecko$. The single spectrum spanned 60 \AA\/ and was centered on 3947 \AA\/. A representative spectrum of HD~179949 is shown in Figure 1. Five nights were allocated in each of three semesters: August 2001, July 2002 and August 2002.  Poor weather and technical issues limited usable data to $\sim$ 10 nights.  Spectra with comparable S/N were taken of two stars known not to have close-in giant planets, $\tau$ Ceti and the Sun (sky spectra were taken at dusk).
  	
The data reduction consisted of the subtraction of the appropriate darks from the stellar, arc and flat-field exposures rather than using the convential biases to remove the baseline from each observation.  The flats were combined and normalized to a mean value of 1 along each row of the dispersion axis.  All the stellar exposures for each night were averaged in order to define a single aperture to use for the spectral extraction. This included a subtraction of the residual background level between orders. The aperture was used to extract 1-dimensional spectra from the stellar, comparison and flat-field exposures. A mean, normalized 1-dimensional flat was divided into the extracted spectra to obtain the most consistent flat-fielded spectra possible.  The wavelength calibration was done with Thorium-Argon arcs. Heliocentric and differential radial velocity corrections were applied. 

In 2001 Th-Ar arcs were taken with every target while in the 2002 observing runs, each stellar exposure was bracketed with an arc. This allowed differential radial velocities ($\Delta$$RV$) to be measured to better than 20 m/s. Detailed discussion on the $RV$ can be found in Walker et al. (2003). Current orbital ephemerides and hence accurate phases ($\pm 0.03$) were determined for each observation by fitting a least-squares sine function with the known radial velocity amplitude and orbital period.  A pure sinusoid was sufficient since all five systems have eccentricity $e \leq 0.05$ (Schneider 2003). Table 2 contains the July 2002 times of sub-planetary position along with the revised periods.

For each star, portions of the spectra, each 7 \AA\/ wide, were extracted 
centered on Ca~II H, K and the strong Al~I line at 3944 \AA.  As a means of normalizing each sub-spectrum, the continuum was set to 1 at the ends and fit with a straight line.  The spectra were grouped by date and a nightly mean was computed for each of the three lines.  An overall average was then taken over all the nights from which we measured nightly residuals.  Each residual spectrum also had a broad, low-order curvature removed.


\begin{deluxetable}{ccccccccl}
\tabletypesize{\footnotesize}
\tablecaption{ The Program Stars \label{table} }
\tablewidth{0pt}
\tablehead{   
\colhead{~ Star} & 
\colhead{Spectral Type} & 
\colhead{$v$sin$i$\tablenotemark{a}} &
\colhead{$M_{p}sini$\tablenotemark{b}} &
\colhead{$P_{orb}$\tablenotemark{b}} &
\colhead{Semi-major axis\tablenotemark{b}} &
\colhead{Ca II K Flux \tablenotemark{c}} 
\\
\colhead{} &
\colhead{} &
\colhead{km s$^{-1}$ } &
\colhead{$M_{J}$} &
\colhead{days} &
\colhead{AU} &
\colhead{\AA} &

}
\startdata
$\tau$ Boo & F7 VI &14.8 $\pm$ 0.3&3.87&3.3128&0.046&0.326\\
HD 179949&F8 V&6.3 $\pm$ 0.9&0.84&3.093&0.045&0.358\\ 
HD 209458&G0 V&4.2 $\pm$ 0.5&0.69\tablenotemark{d}&3.524738&0.045&0.192\\
51 Peg&G2 IV&2.4 $\pm$ 0.3&0.47&4.2293&0.05&0.177\\
$\upsilon$ And&F7 V&9.0 $\pm$ 0.4&0.71\tablenotemark{e}&4.617&0.059&0.252\\
$\tau$ Cet&G8 V& $<$ 2&$-$&$-$&$-$&0.201\\Sun&G2 V&1.73 $\pm$ 0.3&$-$&$-$&$-$&0.300\\ 
\enddata

\tablenotetext{a}{$v$sin$i$ references: $\tau$~Boo (Gray 1982), HD~179949 (Groot et al. 1996), HD~209458 (Mazeh et al. 2000), 51~Peg and Sun (Fran\c{c}ois et al. 1996), $\upsilon$ And (Gray 1986), $\tau$~Ceti (Fekel 1997)}
\tablenotetext{b}{published orbital solutions: $\tau$ Boo \& $\upsilon$ And (Butler et al. 1997), HD 179949 (Tinney et al. 2000), 51 Peg (Marcy et al. 1996), HD 209458 (Charbonneau et al. 1999)}
\tablenotetext{c}{total integrated flux of mean normalized K core}
\tablenotetext{d}{transiting system; $i = 86.1^{\circ}\pm 0.1$ (Mazeh et al. 2000)}
\tablenotetext{e}{closest of three known planets in the system}

\end{deluxetable}



\begin{deluxetable}{ccccccl}
\tabletypesize{\footnotesize}
\tablecaption{ Ephemerides, July 2002 \label{table} }
\tablewidth{0pt}
\tablehead{   
\colhead{~ Star} & 
\colhead{HJD at $\phi=0$} &
\colhead{$\delta$(HJD)\tablenotemark{a}} &
\colhead{Revised $P_{orb}$} & 
\colhead{$\delta(P_{orb})$\tablenotemark{a}} &
\\
\colhead{} &
\colhead{days} &
\colhead{days} &
\colhead{days} &
\colhead{days} &

}
\startdata
$\tau$ Boo & 2452478.770 & 0.099 & 3.31245 & 0.00033\\
HD 179949 & 2452479.823 & 0.093 & 3.09285 & 0.00056\\ 
HD 209458 & 2452481.129 & 0.106 & 3.52443 & 0.00045\\
51 Peg & 2452481.108 & 0.127 & 4.23067 & 0.00024\\
$\upsilon$ And & 2452481.889 & 0.139 & 4.61794 & 0.00064\\
\enddata

\tablenotetext{a}{uncertainties in the respective measurements}
\end{deluxetable}


\section{Results}

The program stars were originally chosen for Doppler planet searches because of their weak chromospheric activity as indicated by the modest Ca II H \& K emission flux measured by the Mt. Wilson $S_{HK}$ index.  However there are marked differences in the reversal structure of each of these targets caused by their different rotational rates (which produce a range of emission strengths and Doppler broadening) as well as by other intrinsic mechanisms, presumably acoustic waves which depend on $T_{eff}$ and $g$.    
The integrated flux of the mean, normalized Ca~II~K core, including the photospheric contribution and the chromospheric emission, is listed in Table 1 and plotted against $v$sin$i$ in Figure 2.  Even with the small number of program stars, there is a reasonable linear relationship between the two parameters as reported in Pasquini et al. (2000).  HD~179949 lies significantly above the best-fit line.  The Sun does as well, but this is known to be an effect of increased activity observed in the 2002 observing runs.\footnote{We observed the naked-eye sunspot grouping of July 2002 in transit during our CFHT run with the Sunspotter$^{\circledR}$ Solar Telescope supplied by the Visitor Information Center at the Onizuka Center for International Astronomy.}    

The residuals of the normalized spectra (smoothed by 21 pixels) taken from the mean were used to generate the Mean Absolute Deviation (MAD = $N^{-1}\Sigma|data_{i}-mean|$ for $N$ spectra) plot shown in Figure 3 (top) for HD~179949.  Superimposed on the plot is the corresponding K-core. The nightly residuals used to generate the MAD plot for HD~179949 are displayed just below. The activity seen in four of the five stars with planets, HD~179949, $\tau$~Boo, $\upsilon$~And and HD 209458, show significant deviations in the H \& K reversals.  The widths of the MAD plots (Figure 4) are the same as those of the reversals themselves as illustrated in Figure 3 for HD~179949. In the Sun's case, images from the Michelson Doppler Imager (MDI) aboard the Solar Heliospheric Observatory (SOHO)\footnote{http://sohowww.nascom.nasa.gov/ and http://soi.stanford.edu/} confirm that the sunspot group was receding which caused the apparent asymmetry of the solar MAD plot. The strong coherence between the H and K activity can be seen in Figure 5 where the integrated mean absolute deviation for H and K is plotted for each star.  The strong aluminum line shows no comparable variations. For all the program stars, the mean absolute deviation of the Al~I line at 3944\AA\/ is less than 0.0005 of the normalized mean.  This indicates that the photospheres of the stars are stable to this level providing additional evidence that the variations seen in the Ca II H \& K reversals are not caused by upper-level photospheric changes but rather are confined to the chromosphere.

\section{HD 179949: A Convincing Case of Planetary-induced Modulation}

HD 179949 shows the greatest nightly modulation of all the stars in all three observing runs.  The integrated flux of the residuals shown in Figure 3 (bottom) are plotted against orbital phase in Figure 6.  The emission clearly increases by $\sim$ 4\% when the planet is in front of the star, at phases near the sub-planetary point ($\phi \sim 0$) and is least when the planet is behind the star ($\phi \sim$ 0.5).  The amplitude and period of this activity has persisted during the year between observations (equal to 108 orbits or at least 37 stellar rotations). A periodogram of even these few data points (Stellingwerf 1978) produces the most significant peak with a period of 3 days.

\subsection{Rotation of HD 179949}

HD 179949 has no photometrically measured rotation period so, as derived from the $v$sin$i$, $P_{rot}\leq$ 9 days. The solid line in Figure 6 is a truncated, best-fit sine curve with $P = P_{orb}$ corresponding to the change in projected area of a bright spot on the stellar surface before being occulted by the stellar limb.  The peak of the model leads the sub-planetary point by 0.17 in phase. If this effect is real, it could provide a constraint on future models of the interaction. Since Ca~II emission is optically thin outside the self-absorption core, limb brightening needs to be included in the model.  According to solar observations discussed by Engvold (1966), limb brightening would increase emission by 3\%, essentially not changing the geometric model.  

The best-fit spot model requires the bright spot to be at a latitude of 
$30^{\circ}$ and a stellar inclination angle $i$ of $87^{\circ}$. Since 
there were no transits detected for this system, an upper limit for the 
orbital inclination is set at $83^{\circ}$ (Tinney et al. 2001). The 
dashed model in Figure 6 has $i = 83^{\circ}$ as it is generally assumed 
that the orbital and stellar inclination angles are comparable. These 
models are inconsistent with the possibility that the star may be tidally 
synchronized with the planet's orbit ($P_{rot}=P_{orb}$). However, from 
Figure 2, the Ca~II~K emission in HD~179949 is unusually strong compared 
to the other targets.  This implies that either the inclination of the 
star is low, $\sim 21^{\circ}$ ($v \simeq$ 17.5 km s$^{-1}$  and 
$P_{rot} \approx P_{orb} \approx$ 3 d.) or that $P_{rot}$ is closer to 9 days and the Ca~II
emission is significantly enhanced in HD~179949 relative to the other stars.

If the star is indeed tidally synchonized with the planet, as is
believed to be the case for $\tau$~Boo, then the modulation in the
emission observed with the 3.093-day period may be tracking a persistent
stellar surface feature.  However, tidal synchronization of HD~179949 is
not a favoured scenario for two reasons.  Firstly, the tidal
synchonization timescale for the system $\approx$ 700 Gyr using a
$M_{p}$sin$i = 0.84 M_{J}$. (See Equation 1 of Drake et al. 1998).  
Even if $i = 21^{\circ}$, as would be the case for  $v$ = 17.5 km s$^{-1}$ and 
$M_{p} = 2.34 M_{J}$, then the sychronization timescale is still very long at 79 Gyr.  Secondly,
ROSAT X-ray data list HD~179949 as having at least double the X-ray flux (a measurement independent of $i$) as compared to other single F8-9 dwarfs and $\sim$ 10 times that of the Sun's.  H\"{u}nsch et al. (1998) report
the X-ray flux for HD~179949 to be $L_{X}$ = 41.0 $\times 10^{27}$ erg/s while for 36~UMa, a nearly identical star,  $L_{X}$ = 18.4 $\times 10^{27}$ erg/s .

\section{Summary}

From the sample of stars observed, those with planets (with the exception of 51~Peg) show significant nightly variation in their Ca~II H \& K cores. $\tau$~Ceti which has no close planet remained very steady over all three observing runs. The Sun's Ca~II emission  was stable within each run. However, a large naked-eye sunspot grouping appeared in the summer of 2002 and was detected as an enhancement in the H \& K reversals accounting for its MAD signal (Figure 4).  There is enhanced nightly activity in the stars with close-in giant planets whose variations may be cyclical and phase shifted. Further observations of these systems at varying orbital phases are required. 

In the one clear case, HD~179949 exhibits repeated phase-dependent activity with enhanced emission near the sub-planetary point and less emission half an orbit later.  This is consistent with a magnetic heating scenario and may be a first indirect glimpse at a magnetosphere of an extrasolar planet. Additional Ca II observations are crucial to confirm the continuity of the magnetic interaction as well as to establish better phase coverage.  Due to the nearly exact 3-day orbit, it is impossible to get many distinct phases during one observing season from the ground. Space based observations will not only increase the phase coverage, but will also allow observations of other activity diagnostics such as transition region and coronal lines in the UV and FUV. Such mapping of planet-induced activity at varying stellar atmospheric depths will allow characterization and quantification of the physical interaction between the magnetically heated layers of the star and the planet's magnetosphere.  

\acknowledgements

Research funding from the Canadian Natural Sciences and Engineering
Research Council (G.A.H.W. \& E.S.) and the National Research Council
of Canada (D.A.B.) is gratefully acknowledged. We are also indebted to the
CFHT staff for their care in setting up the CAFE fiber feed and the
Gecko spectrograph.

\clearpage

\begin{figure}
\plotone{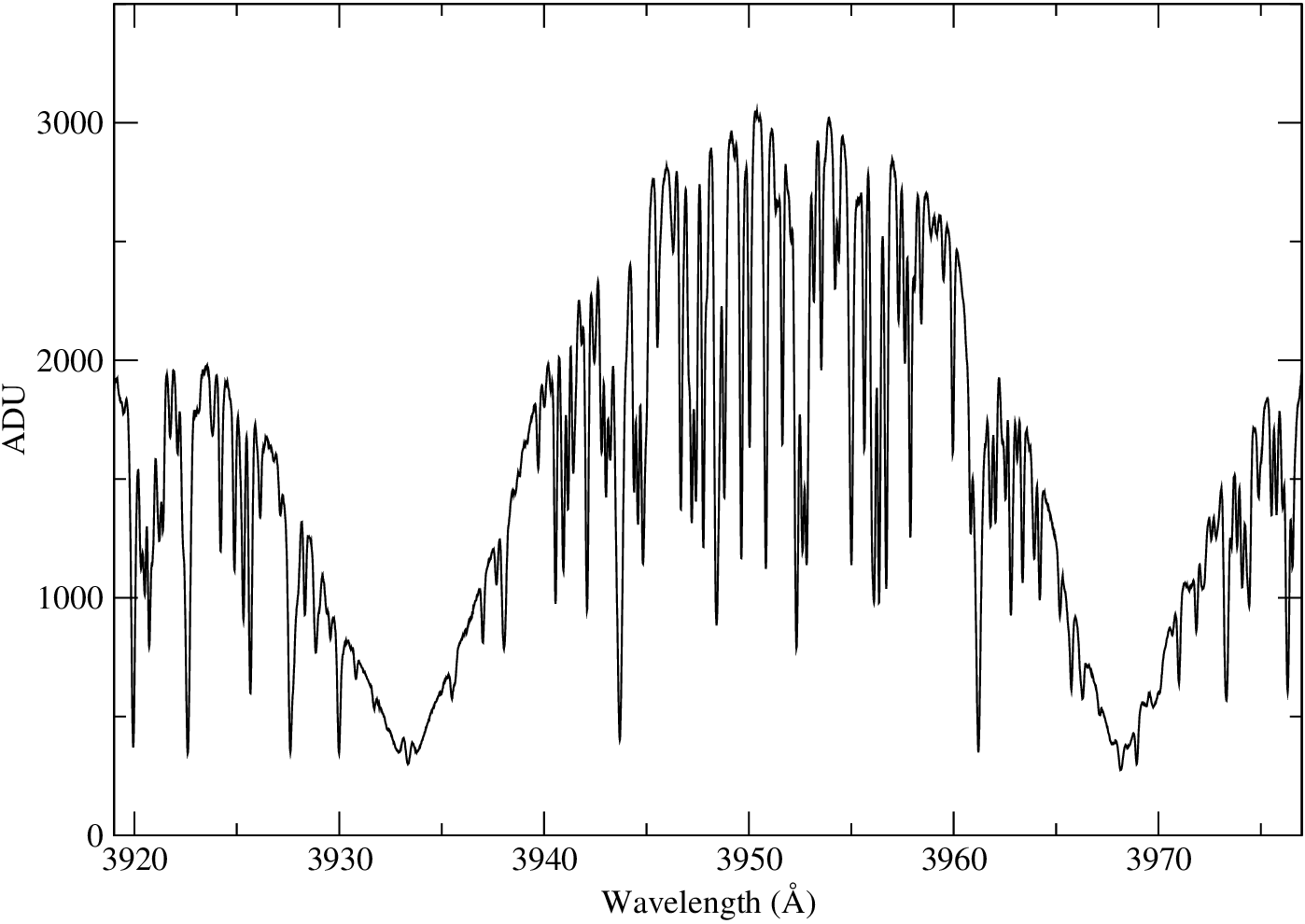}
\caption{A single flat-fielded spectrum of HD~179949 taken over 40 minutes of integration time.\label{fig1}}
\end{figure}

\begin{figure}
\plotone{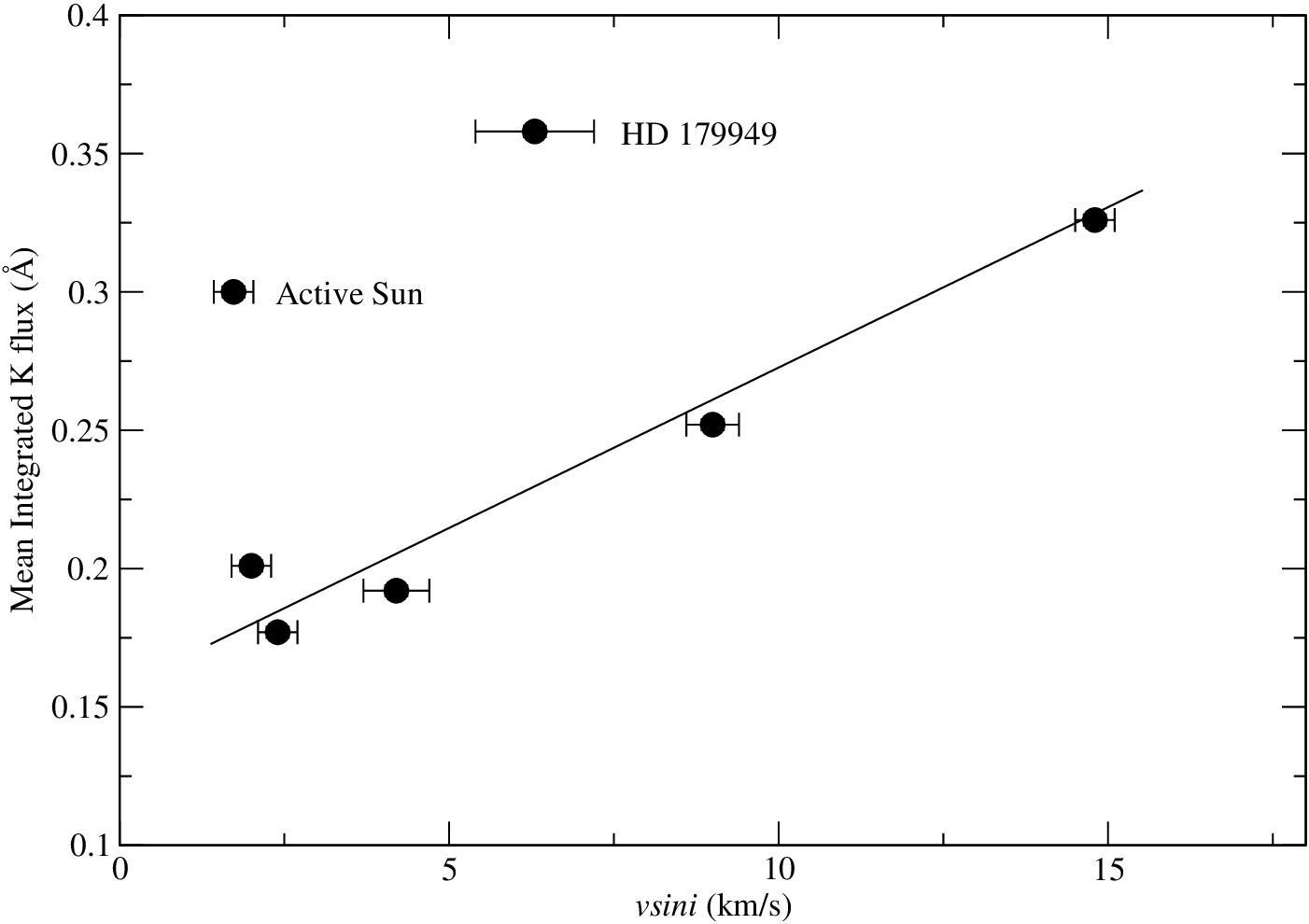}
\caption{Published $v$sin$i$ values for the program stars (see Table 1) plotted against the integrated flux of the mean normalized Ca II K cores.\label{fig2}}
\end{figure}

\begin{figure}[h]
\epsscale{0.85}
\plotone{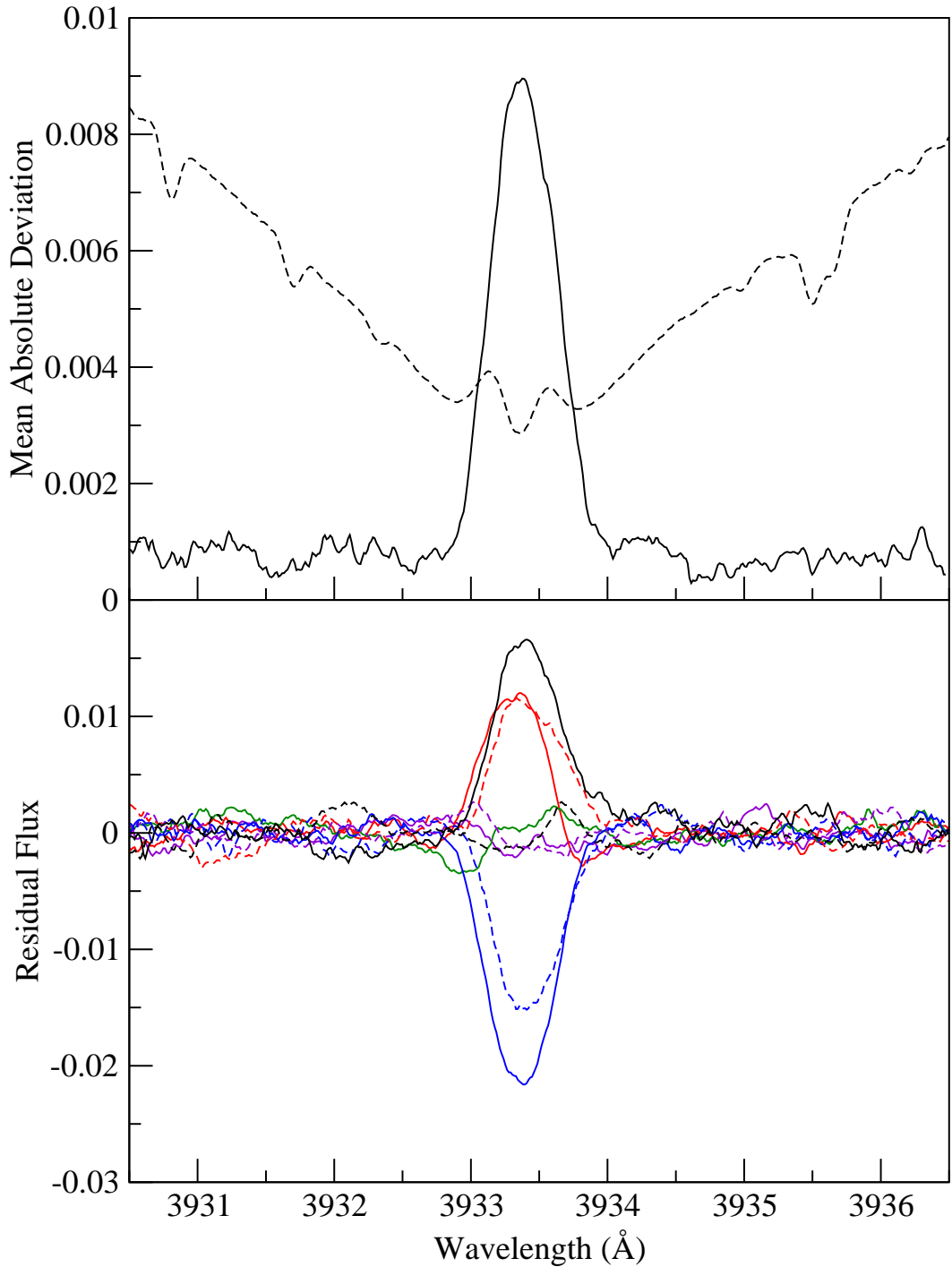}
\caption{Top: The mean absolute deviation (MAD) of the Ca II K core (solid line) of HD~179949. The units are intensity as a fraction of the continuum. Overlayed (dashed line) is the overall mean spectrum indicating that the activity in HD~179949 is confined to the K emission. Bottom: Residuals (smoothed by 21 pixels) from the normalized mean spectrum of the Ca~II~K core of HD~179949.  The residuals for the H line follow these closely.  \label{fig3}}
\end{figure}


\begin{figure}[h]
\epsscale{0.85}
\plotone{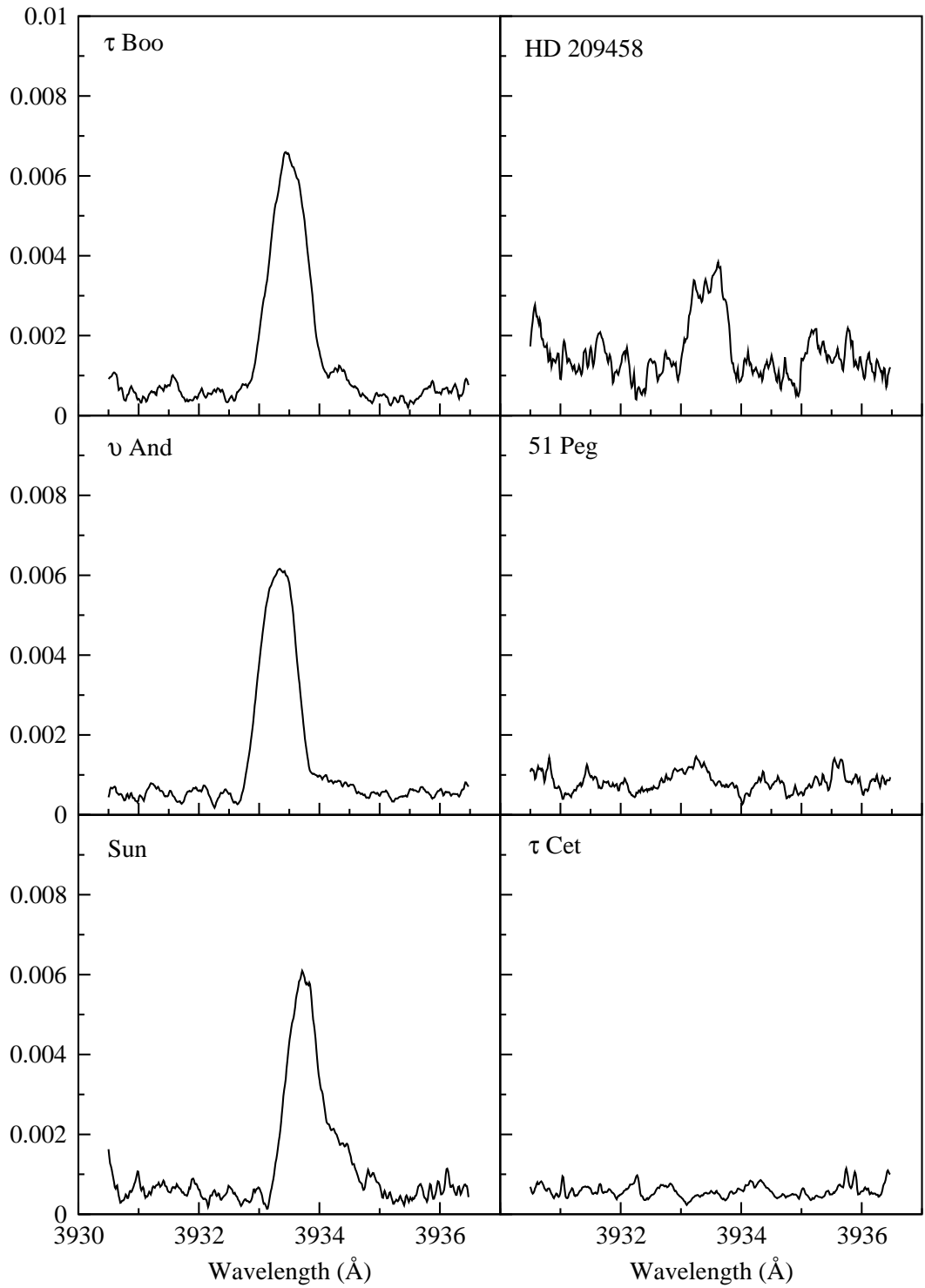}
\caption{MAD plots for the other six program stars. Units are intensity as a fraction of the continuum.\label{fig4}}
\end{figure}

\begin{figure}
\plotone{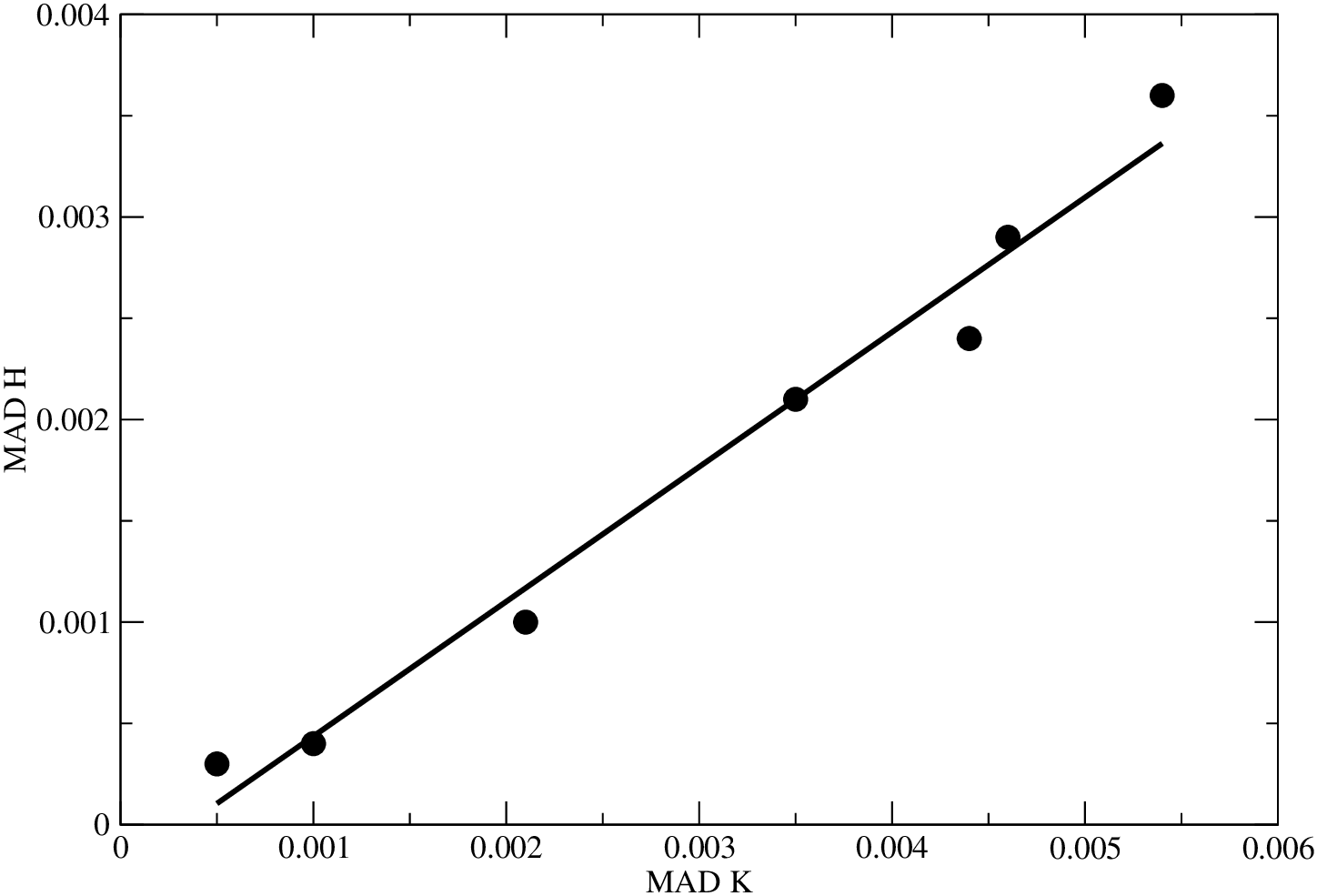}
\caption{Integrated ``intensity'' of the MAD H and K plots for all the program stars.  The slope of the best-fit line is 0.665.}
\end{figure}

\begin{figure}
\plotone{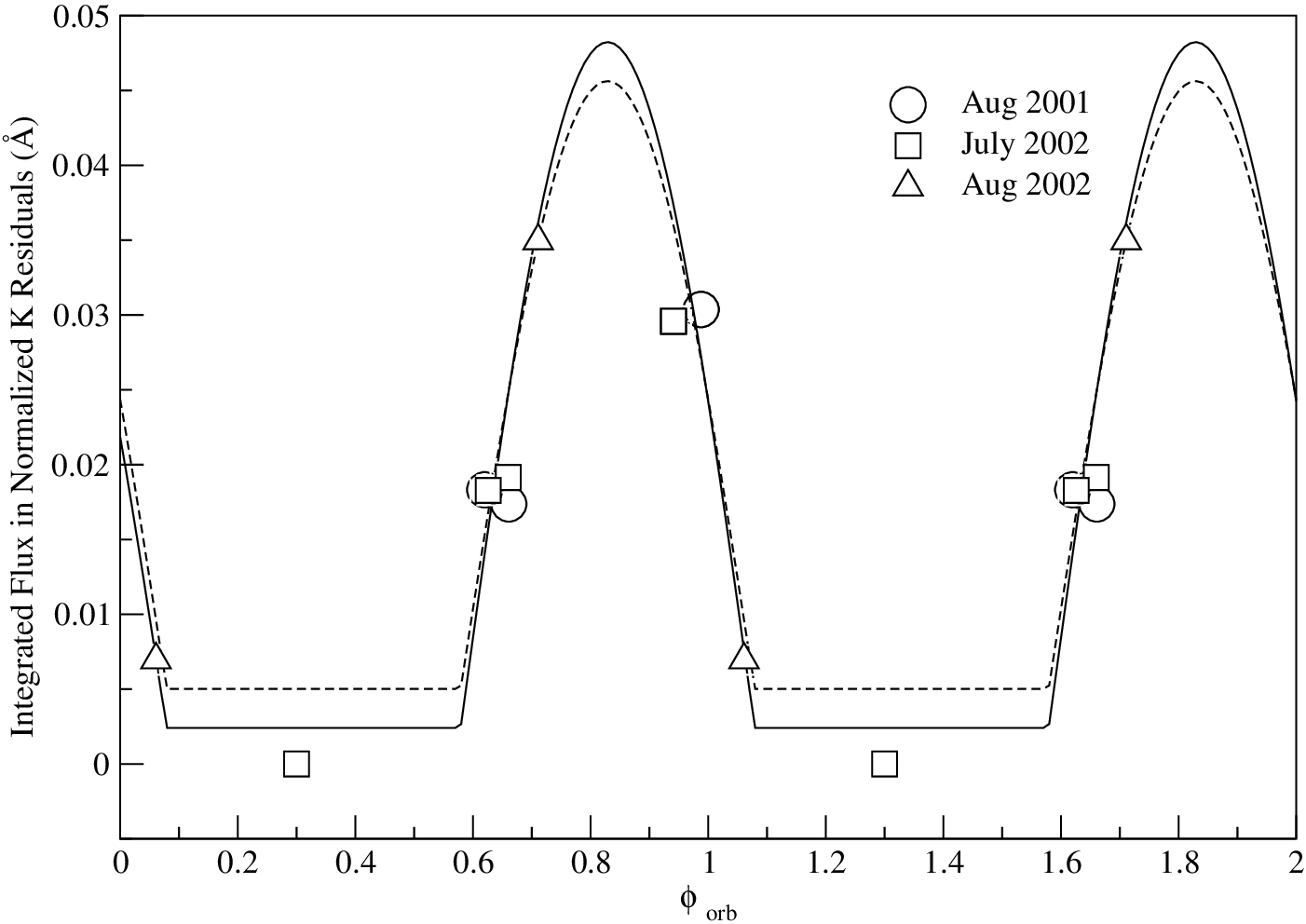}
\caption{Integrated flux of the nine K-line residuals taken from a normalized mean spectrum.  The minimum flux was set to zero and all others scaled accordingly. The size of the points is the size of the error bars.
The solid line is the best-fit bright-spot model discussed in the text with the spot at a latitude of $30^{\circ}$ and stellar inclination angle $i = 87^{\circ}$. The dashed line is a model with $i = 83^{\circ}$}.
\end{figure}

\end{document}